\documentstyle[12pt,epsf]{article}
\newcommand{\vs}{\vspace{0.25cm}}

\newtheorem{theorem}{Theorem}
\newtheorem{itlemma}{Lemma}[section]
\newtheorem{itproposition}[itlemma]{Proposition}
\newtheorem{itcorollary}[itlemma]{Corollary}
\newtheorem{itremark}[itlemma]{Remark}
\newtheorem{itremarks}[itlemma]{Remarks}
\newtheorem{itdefinition}[itlemma]{Definition}
\newtheorem{itexample}[itlemma]{Example}

\newenvironment{lemma}{\begin{itlemma}\rm}{\end{itlemma}} %no-italics
\newenvironment{remark}{\begin{itremark}\rm}{\end{itremark}} %no-italics
\newenvironment{remarks}{\begin{itremarks} \rm}{\end{itremarks}}
\newenvironment{corollary}{\begin{itcorollary}\rm}{\end{itcorollary}}
\newenvironment{proposition}{\begin{itproposition}\rm}{\end{itproposition}}
\newenvironment{definition}{\begin{itdefinition}\rm}{\end{itdefinition}}
\newenvironment{example}{\begin{itexample}\rm}{\end{itexample}}
\newenvironment{fact}{\noindent {\em Fact}. \ \ }{\hfill \medskip}
\newenvironment{proof}{\noindent {\em Proof}.\ \
}{\hspace*{\fill}$\Box$\medskip}
\newenvironment{claim}{\noindent {\em Claim}. \ \ }{\hfill \medskip}
\newcommand{\be}[1]{\begin{equation}\label{#1}}
\newcommand{\ee}{\end{equation}}
\newcommand{\bl}[1]{\begin{lemma}\label{#1}}
\newcommand{\br}[1]{\begin{remark}\label{#1}}
\newcommand{\brs}[1]{\begin{remarks}\label{#1}}
\newcommand{\bt}[1]{\begin{theorem}\label{#1}}
\newcommand{\bd}[1]{\begin{definition}\label{#1}}
\newcommand{\bp}[1]{\begin{proposition}\label{#1}}
\newcommand{\bc}[1]{\begin{corollary}\label{#1}}
\newcommand{\bfact}[1]{\begin{fact}\label{#1}}
\newcommand{\bex}[1]{\begin{example}\label{#1}}
\newcommand{\ec}{\end{corollary}}
\newcommand{\efact}{\end{fact}}
\newcommand{\eex}{\end{example}}
\newcommand{\el}{\end{lemma}}
\newcommand{\er}{\end{remark}}
\newcommand{\ers}{\end{remarks}}
\newcommand{\et}{\end{theorem}}
\newcommand{\ed}{\end{definition}}
\newcommand{\ep}{\end{proposition}}
\newcommand{\epr}{\end{proof}}
\newcommand{\bpr}{\begin{proof}}
\newcommand{\bcl}{\begin{claim}}
\newcommand{\ecl}{\end{claim}}

\newcommand{\bi}{\begin{itemize}}
\newcommand{\ei}{\end{itemize}}
\newcommand{\ben}{\begin{enumerate}}
\newcommand{\een}{\end{enumerate}}
\newcommand{\text}[1]{\hbox{\rm \ #1\ \/}}

\title{Connection Between Continuous and Discrete Time
Quantum Walks on  $d$-Dimensional Lattices; Extensions to General
Graphs}

\vs

\vs

\author{Domenico D'Alessandro\thanks{Department of Mathematics, Iowa State
University, Ames, Iowa, U.S.A.\ \ Electronic address:
daless@iastate.edu}}

\begin{document}

\maketitle

 \vs

\begin{abstract}
%abstract

I obtain the dynamics of the continuous time quantum walk on a
$d$-dimensional lattice, with periodic boundary conditions, as an
appropriate limit of the dynamics of the discrete time quantum walk
on the same lattice. This extends the main  result of \cite{Strauch}
which proved this limit for the case of the quantum walk on the
infinite line  and the quantum cellular automaton proposed in
\cite{BB}. By highlighting the main features of the limiting
procedure, I then extend it to general graphs. For a given discrete
time quantum walk on a general graph, I single out  the type   of
continuous dynamics (Hamiltonians) that can be obtained as a limit
of the discrete time  dynamics.

\end{abstract}

\section{Generalities on quantum walks; continuous and discrete time}
\label{intro}

Consider a graph $G:=\{V, E \} $ with a set of vertices $V$ of
cardinality $N$  and a set of edges $E$. We assume $G$ to be
undirected and without self-loops.  In the interval of time $\Delta
t$ a certain fraction $\gamma \Delta t$ of a quantity $p_j$ leaves
the location of the $j$-th vertex to move to a neighboring vertex
$k$. The quantity $p_j$, $j=1,\ldots,N$,  at time $t+\Delta t$ is
given by\footnote{The rate $\gamma$ may  depend on time and we could
have allowed it to depend on the location $j$. In the latter case
though we cannot choose a  Laplacian matrix $L$ (see later)  to be
symmetric.} \be{pjtdt} p_j(t+\Delta t) =p_j(t)-\text{deg}(j) \gamma
\Delta t p_j(t) + \sum_{(k,j) \in E} \gamma \Delta t p_k(t). \ee
Define $\vec p$ the $N-$vector whose components are the $p_j$'s
quantities, and $L$ the {\it Laplacian} matrix defined by
$L_{jj}=-\text{deg}{(j)}$, $L_{jk}=1$ if $(j,k) \in E$ and
$L_{jk}=0$ otherwise. $L$ is symmetric and, except for the elements
on the diagonal, coincides with the adjacency matrix $A$, that is,
$A_{jk}:=L_{jk}$, when $j \not=k$ and $A_{jk}=0$ if $j=k$. Taking
the limit $\Delta t \rightarrow 0$ in (\ref{pjtdt}) one obtains the
differential equation for the {\it classical continuous-time  random
walk} (CTRW). \be{clasRW} \frac{d}{dt} \vec p=\gamma L \vec p. \ee
The reason to call this model a random walk is that the $p_j(t)$'s
may represent the probability of a walker to be in position $j$ at
time $t$. A continuous-time {\it quantum random walk} (CTQW) (see
\cite{Farhi}) is obtained by quantizing equation (\ref{clasRW}). One
replaces $\vec p$ with a complex $N$-vector representing the state
of a quantum system, while $L$ is taken as the Hamiltonian
determining the evolution according to the Schr\"odinger equation
\be{Scrod} i \dot \psi=\gamma L\psi,  \ee with $|\psi_j|^2$
representing  the probability of the quantum system being in the
basis state $|j \rangle $. This definition can be generalized by
using  an Hamiltonian  different from $L$ which respects the
topology of the graph $G$, that is elements different from zero
correspond to edges in the graph. For example one can use the
adjacency matrix $A$ instead of $L$ in (\ref{Scrod}).\footnote{In
this case, if $G$ is a regular graph, i.e., deg($j$) is the same for
every $j$ in $V$, the corresponding dynamics would differ from the
ones of (\ref{Scrod}) only by a physically unimportant phase
factor.}

The {\it classical discrete-time random walk} (DTRW) is obtained by
discretizing equation (\ref{clasRW}) and therefore  it is given by
equation (\ref{pjtdt}). The CTRW is obtained from the DTRW by taking
the limit $\Delta t \rightarrow 0$ in equation (\ref{pjtdt}) as seen
above. To define the {\it discrete time quantum walk} (DTQW), we
would like to write an equation of the form \be{forDTQW} \psi(t+1)=U
\psi(t), \ee with $\psi$ the state of a quantum system and  $U$  a
unitary operator representing a (closed) quantum evolution.  We also
 would like to have a $U$ which respects the structure of the
underlying graph $G$, i.e., $U_{jk}$, with $j \not= k$  is different
from zero if and only if there exists an edge $(j,k)$ in $E$.
Unfortunately, only special graphs have the property that such a
unitary matrix exists \cite{Simoseve}. In order to give a definition
which is suitable for any graph one proceeds as follows.

Let ${\cal V}$ be the Hilbert space spanned by the orthonormal
states $\{ |j \rangle \}$, with $j \in V$. Let ${\cal E}$ be the
subspace of ${\cal V} \otimes {\cal V}$ spanned by  $|j,k\rangle$
with $(j,k) \in E$. The basis state $|j,k\rangle $ represents the
state of a walker which is currently in vertex $j$ and is moving to
vertex $k$, i.e., $j$ and $k$ are the {\it present} and {\it future}
location of the walker, respectively. On the space $\cal E$, the
evolution $U$ of the DTQW is of the form $ U = W \tilde C$. The
operation $\tilde C$, called {\it coin tossing}, is of the form
\be{CT} \tilde C= \sum_{j \in V} |j \rangle \langle j| \otimes Q_j,
\ee where $Q_j$ is a unitary transformation on ${\cal V}$, depending
on $j$. The subspace spanned by the states corresponding to the
neighboring vertices of $j$ (and therefore its orthogonal
complement) is invariant under $Q_j$. The operation $W$ is any
unitary which transforms the elements $|j,k\rangle$ as $W|j,k\rangle
=|k,r\rangle$, for some $|k,r\rangle \in {\cal E}$, i.e, it moves
the future state in the present state position. One possibility is
the {\it swap} operation, defined by $W |j,k \rangle =|k,j \rangle$,
since $|j,k\rangle \in {\cal E} \leftrightarrow |k,j\rangle \in
{\cal E}$.

If the graph $G$ is regular, one can give a definition which makes
the coin's role   more transparent. We call this quantum walk the
{\it coined} DTQW. It is defined as follows. Let $m$ be the degree
of $G$ and consider a (coin) space ${\cal C}$ spanned by orthogonal
states $\{|c_1\rangle, \ldots, |c_m \rangle\}$ each representing the
result of a coin tossing. Denote by $n_j(c_k)$, $j=1,\ldots,N$,
$k=1,\ldots,m$, an element in $V$ if a coin result $c_k$ induces a
transition from $j$ to it. The coined DTQW evolves on ${\cal C}
\otimes {\cal V}$ as $R=S C \otimes {\bf 1}$, where $C$ is a unitary
(coin tossing) operation on ${\cal C}$, ${\bf 1}$ is the identity on
the walker space  ${\cal V}$ and $S$ is a {\it controlled shift}
defined by \be{defcontrS} S|c_k, j \rangle =|c_k, n_j(c_k) \rangle.
\ee The coined DTQW, defined in the case of a regular graph, is a
special case of the more general DTQW defined above, when the coin
operations $Q_j$ in (\ref{CT}) are essentially independent of $j$
and $W$ takes a special form. To be more specific, let $C$ in the
coined DTQW be defined  by \be{defC} C|c_k\rangle := \sum_{l=1}^m
\alpha_{lk} |c_l \rangle.  \ee Then choose  $Q_j$ in (\ref{CT}) as
\be{Qj} Q_j|n_j(c_k)\rangle=\sum_{l=1}^m \alpha_{lk} |n_j(c_l)
\rangle \ee (notice the coefficients $\alpha_{lk}$ are independent
of $j$). The transformation $W$ of the DTQW is chosen as \be{W}
W|j,n_j(c_k) \rangle = |n_j(c_k), n_{n_j(c_k)}(c_k)\rangle.  \ee
With these choices, the dynamics of the DTQW on ${\cal E}$ are
equivalent to the dynamics of the coined DTQW on ${\cal C} \otimes
{\cal V}$. In fact, using the isomorphism $\chi :  {\cal C} \otimes
{\cal V} \rightarrow {\cal E}$ defined by \be{defisT} \chi |c_k,j
\rangle = |j,n_j(c_k) \rangle, \ee we have that $ \chi S(C \otimes
{\bf 1})=W \tilde C \chi $. In fact, a straightforward computation
shows \be{pod} \chi S(C\otimes {\bf 1}) |c_k,j \rangle =
\sum_{l=1}^m \alpha_{lk} |n_j(c_l), n_{n_j(c_l)}(c_l) \rangle= W
\tilde C \chi |c_k,j \rangle. \ee

\vs

Having defined the DTQW and the CTQW the question remains on whether
the CTQW can be obtained as a limit of the DTQW on the same graph.
This is a fundamental question which was posed in \cite{Kemperev}
and  has received much attention recently (see, e.g., \cite{Childs},
\cite{Strauch} and references therein).  It is motivated, among
other things, by the fact that CTQW and DTQW on the same graphs have
showed similar behavior in several applications. Since the DTQW
evolves on an higher dimensional space as compared with the CTQW,
the procedure to obtain the CTQW from the DTQW should involve not
only an appropriate limit but also some sort of a projection of the
dynamics onto a lower dimensional space. This work was done in
\cite{Strauch} for the quantum walk on the infinite line. In this
paper, I extend and somewhat simplify the calculation in
\cite{Strauch} to prove the same limit for the coined DTQW on a
$d$-dimensional lattice with periodic boundary conditions. In
particular for $d=1$, I obtain the CTQW from the DTQW for the cycle.
I then observe that this generalization is only an example of a
general procedure that I then extend to general graphs. I formalize
this procedure by describing the set of Hamiltonian dynamics for the
CTQW  which can be obtained by a limit of the dynamics of the DTQW
(cf. Theorem \ref{CARA}). In the following, I shall consider only
coined DTQW.

The paper is organized as follows. In Section \ref{lattici} I
describe in detail the quantum walk on a $d$-lattice. The
$d$-lattice is the Cartesian product of $d$ cycle graphs. If the
transition from discrete to continuous is obtained for each factor
of a product graph then it carries over to the whole graph (cf. end
of Section \ref{lattici} and Remark \ref{noproblem}). Therefore, it
is enough to restrict ourselves to the cycle. I prove the transition
from DTQW to CTQW for the cycle in Section \ref{Main}. The
calculation in this section can be carried over to other quantum
walks. Moreover, even for the case of the cycle alternative limiting
procedures to obtain the CTQW from the DTQW can be devised as
discussed in Remark \ref{evencyc}. In Section \ref{extension}, I
extend the procedure to general coined quantum walks. I characterize
in Theorem \ref{CARA} the set of Hamiltonians on the space ${\cal E}
\otimes {\cal V}$ whose dynamics can be obtained as a limit of the
one of the DTQW. I give an example in section \ref{Example} and
conclude in section \ref{conclu}.

\section{Quantum walks on $d$-dimensional lattices}
\label{lattici}

Consider a coined DTQW on  a $d$-dimensional lattice whose set of
vertices $V$ is given by $V=\{0,1,\ldots,N-1\}^d$. If $\tilde {\cal
V}:=\text{span} \{ |0 \rangle, |1\rangle, \ldots, |N-1 \rangle \}$,
is  the  Hilbert space associated with $\{ 0,1, \ldots, N-1\}$, the
walker Hilbert space for this graph is  ${\cal V}= \otimes^d \tilde
{\cal V}$, which is spanned by the orthonormal  vectors
$|j_1,\ldots, j_d\rangle$, $j_1,\ldots,j_d=0,1,,\ldots,N-1$, which
represent  vertices  of the graph with coordinates $j_1,\ldots,j_d$.
In the graph $G$, the vertex labeled by $(j_1,\ldots,j_d)$ has $2d$
neighbors each differing by $(j_1,\ldots,j_d)$ by only one
coordinate and with (Hamming) distance $1$, i.e., of the form
$(j_1,\ldots,j_k \pm 1, \ldots,j_d)$, where the $\pm 1$ operation
has to be intended $\text{mod} N$. Let $\tilde G$ be the graph
representing the cycle with $N$ nodes. $G$ is the product of $d$
copies of $\tilde G$.  Its adjacency matrix is \be{adjacency}
A:=\sum_{l=1}^d \tilde A^{(l)}, \ee where $\tilde A^{(l)}$ is the
tensor product of $d$, $N \times N$ identity matrices except in the
$l$-th position which is occupied by the adjacency matrix of the
cycle $\tilde A$.\footnote{There are several definitions of product
of d graphs. Here we intend  the {\it Cartesian product} of $d$
graphs $G_l\equiv (V_l,E_l)$, $l=1,\ldots,d$ which is defined as the
graph $G\equiv (V,E)$ having the set of vertexes $V$ equal to the
Cartesian product $V:=V_1 \times V_2 \times \ldots \times V_d$ and
set of edges $E$ such that $\left( (u_1,u_2,\ldots, u_d),
(v_1,v_2,\ldots, v_d) \right) \in E$ if and only if all $u_j$'s are
equal to the corresponding $v_j$'s except for exactly one pair
$u_{\bar j} \not= v_{\bar j}$ which is such that $(u_{\bar j},
v_{\bar j}) \in E_{\bar j}$.  The adjacency matrix of such a graph
is $A=\sum_{l=1}^d \tilde A^{(l)}$ where $\tilde A^{(l)}$ where
$\tilde A^{(l)}$ is the tensor product of $d$ matrices all equal to
the identity, with the one in a generic position $r$ of dimension
$|V_r|$, except for the matrix in position $l$ which is equal to the
adjacency matrix of the $l$-th graph. A quick way to see that this
is the case is to recall that a definition of adjacency matrix $A$
given by $e_j^TAe_k=1$ if and only if $(j,k)$ is an edge and $=0$
otherwise. Here $e_j$ and $e_k$ are the standard basis vectors
corresponding the vertexes $j$ and $k$. A quick calculation of
$\langle j_1,\ldots,j_d |A |k_1,\ldots, k_d \rangle$ shows that this
is the case for the above defined matrix. If the graphs $G_1,\ldots,
G_d$ are all regular, so is the product graph and its degree is the
sum of the degrees of the graphs $G_1,\ldots, G_d$.} $\tilde A$ is
given by $\tilde A =F+F^T$, where the circulant matrix $F$ is
defined as \be{circpm} F:=\pmatrix{0 & 0 & \cdot & \cdot & \cdot & 0
& 1 \cr 1 & 0 & \cdot & \cdot & \cdot & 0 & 0 \cr 0 & 1 & \cdot &
\cdot & \cdot & 0 & 0 \cr \cdot  & \cdot & \cdot & \cdot & \cdot &
\cdot & \cdot \cr \cdot & \cdot & \cdot & \cdot & \cdot & \cdot  &
\cdot \cr \cdot & \cdot & \cdot & \cdot & \cdot & \cdot  & \cdot \cr
0 & \cdot & \cdot & \cdot & \cdot & 1 & 0}. \ee The Laplacian is
given by \be{Lapp} L:=-\text{deg}(G) {\bf 1} +A=-2d {\bf 1}+A, \ee
and the solution of the corresponding Schrodinger equation is
$\psi(t)=X(t)\psi(0)$, with $X(t)=X_1(t) \otimes \cdots \otimes
X_d(t)$, where $X_l$ is the solution of the Schr\"odinger operator
equation corresponding to the $l$-th graph (cycle) \be{Scrop} i\dot
X_l= \gamma \tilde L X_l, \qquad X_l(0)={\bf 1}, \ee where $\tilde
L=-2{\bf 1} + \tilde A$ is the Laplacian corresponding to a single
cycle.

\vs

To construct a coined DTQW on the $d$-dimensional lattice, we have
to introduce a coin space ${\cal C}$ which is spanned by $2d$ basis
vectors, each corresponding to a different coin result. We choose
the basis so that the vectors corresponding to the same degree of
freedom are placed one after the other. For example, for a
$2$-dimensional lattice, ${\cal C}$ is spanned by the ordered basis
$\{ | \rightarrow \rangle, | \leftarrow \rangle , |\uparrow \rangle,
|\downarrow \rangle \}$, which induce a right, left, up and down
motion, respectively. The dynamics of the coined DTQW on ${\cal C}
\otimes {\cal V}$ is given by $S C \otimes {\bf 1}_{N^d \times N^d}$
where $C$ is a coin transformation on ${\cal C}$ and $S$ is the
controlled shift. $S$ is a $2dN^d \times 2dN^d$ block diagonal
matrix with $d$, $2 N^d \times 2N^d$-dimensional,  blocks each
corresponding to one degree of freedom in the coin space. The $l$-th
block is given by \be{lthblock} S^{(l)}=\pmatrix{F^{(l)} & 0 \cr 0 &
F^{(l)T}}, \ee where $F^{(l)}$ ($F^{(l)T}$) is the tensor product of
$d$, $N \times N$ identity matrices, except in the $l$-th position
which is occupied by $F$ ($F^T$) in (\ref{circpm}). It represents a
forward (backward) motion in the degree of freedom identified by $l$
(for example $l=1$ (right or left motion), $l=2$ (upward and
downward motion).

The matrix $C$ is, in principle, a general matrix in $U(2d)$ (i.e.,
$2d \times 2d$ unitary), which depends therefore on $(2d)^2$ real
parameters. It is the only element in the dynamics which couples the
various degree of freedom of the motion of the DTQW. We would like
to obtain the dynamics of the CTQW as an appropriate limit involving
the parameters in this matrix and going from discrete to continuous
time. Since the dynamics of the CTQW given by (\ref{Scrop}) is
completely decoupled, it is reasonable to restrict our attention to
coin transformations which do not couple the various degrees of
freedom of the coin. The corresponding matrix $C$ has a block
diagonal structure with $d$, $2 \times 2$ blocks belonging to
$U(2)$. With this restriction, the dynamics of the DTQW also are
completely decoupled and we can restrict our attention to only one
factor in the product graph, i.e., a cycle. Therefore, in the
following, we shall restrict ourselves to a DTQW on a cycle with $N$
vertexes, whose dynamics is given by $S(C \otimes {\bf 1}_{N \times
N})$ with $C$ in $SU(2)$ and \be{esse} S=\pmatrix{F & 0 \cr 0 &
F^T}. \ee

\br{noproblem} In what we have said above, there is no reason to
consider the product of cycles with equal number of  nodes $N$,
other than notational convenience. More generally,  for product
graphs, the adjacency matrix  always has the form of a sum of
adjacency matrices as in (\ref{adjacency}) and the dynamics of the
continuous quantum walks are always decoupled. Therefore if we have
a limiting procedure to obtain the dynamics of the CTQW from the
DTQW for each factor graph, we can can carry this over to the
product graph by restricting the dynamics of the coin space in a way
that the dynamics on the different factor graphs are decoupled. The
same holds for the quantum walk where the Hamiltonian is given by
the Laplacian since this only differs from the one corresponding to
the adjacency matrix by a physically unimportant multiple of the
identity. The important fact is that the Hamiltonian for the product
graph has the form (\ref{adjacency}) (i.e., sum of matrices which
are products of the identity on the various spaces except for a
local Hamiltonian). There are however Hamiltonians that respect the
structure of the product graph but cannot be written this way. For
these Hamiltonians the graph has to be considered as a whole. \er

\section{Continuous time quantum walk as limit of discrete time
quantum walks on the cycle}
\label{Main}

Consider the coined DTQW on the cycle and write the coin operation
$C \in SU(2)$ as \be{coinop} C=Re^{iDx},  \ee with $R:=\pmatrix{0 &
-i \cr -i & 0}$ and $D:=\pmatrix{0 & -1 \cr -1 & 0}$. With $x$
small, $U^2(x)=SC \otimes {\bf 1}SC \otimes {\bf 1}$, represents a
small perturbation to a transformation $U^2(0)=-{\bf 1}_2 \otimes
{\bf 1}_N$ (cf. (\ref{esse}) using the fact that $FF^T={\bf 1}_N$)
which does not modify the state of the system (except for an
unimportant phase factor). Write $e^{iDx}={\bf 1}_2+iDx+O(x^2)$. We
write $U^2(x)$ as\footnote{Sometimes, when it is clear from the
context, we omit the dimension of the identity matrix ${\bf 1}$.}
\be{calcuU2x} U^2(x)=S \left(R({\bf 1}_2+iDx+O(x^2))\otimes {\bf
1}_N \right)  S \left(R({\bf 1}_2+iDx+O(x^2))\otimes {\bf 1}_N
\right)\ee
$$=\left(S(R\otimes {\bf 1}) \right)^2 + i \left( \left( S(R\otimes {\bf
1})\right)^2 D\otimes {\bf 1} + S(R\otimes {\bf 1}) D \otimes {\bf
1} S(R\otimes {\bf 1}) \right)x+ O(x^2). $$ We have already said
that $\left( S(R\otimes {\bf 1})\right)^2=-{\bf 1}_2 \otimes {\bf
1}_N$. On the other hand,  a direct calculation shows that
\be{showd} \left( S(R \otimes {\bf 1})\right)^2 D\otimes {\bf 1} +
S(R \otimes {\bf 1})  D \otimes {\bf 1} S(R\otimes {\bf 1})
=\pmatrix{0 & {\bf 1}+F^2 \cr {\bf 1}+ F^{2T} & 0}. \ee Therefore,
we have \be{lpmko} U^2(x)=-\left( {\bf 1}-i \pmatrix{0 & {\bf 1}+F^2
\cr {\bf 1}+ F^{2T} & 0}x + O(x^2) \right). \ee If $H$ is the
Hamiltonian \be{Hamilt} H:=  \pmatrix{0 & {\bf 1}+F^2 \cr {\bf 1}+
F^{2T} & 0}, \ee then \be{U2x} U^2(x)=-e^{-iHx}+O(x^2), \ee i.e.,
for small perturbations $x$ two iterations of the DTQW give an
evolution which (except for the unimportant phase factor $-1$)
corresponds to the  continuous evolution by a Hamiltonian $H$ in
(\ref{Hamilt}) in the same time $x$. One obtains an evolution $e^{-i
\gamma H t}$ over an arbitrary time $t$ by applying $U^2(x)$ an
infinite number of times and, at the same time, letting $x
\rightarrow 0$. More specifically,  setting $\frac{\tau}{2}=
\frac{\gamma t}{x}$ and neglecting the, physically irrelevant factor
$-1$ in (\ref{U2x}), we have\footnote{This limit can be obtained
from general properties of the logarithms of matrices (cf.
\cite{HornJohnsonT} section 6.5). Set $\frac{\gamma
t}{x}:=m=\frac{\tau}{2}$. The limit in (\ref{limite}) becomes $$
\lim_{m \rightarrow \infty} \left( e^{-i H \frac{\gamma
t}{m}}+O(\frac{1}{m^2}) \right)^m= \lim_{m \rightarrow \infty}
\left({\bf 1}- iH\frac{\gamma t}{m} + O(\frac{1}{m^2}) \right)^m=
\lim_{m \rightarrow \infty} e^{\log \left( {\bf 1} -iH \frac{\gamma
t}{m}+O(\frac{1}{m^2})\right)^m}= $$ \be{limo4} \lim_{m \rightarrow
\infty}e^{-m\left(\frac{1}{m} \left( iH\gamma
t-O(\frac{1}{m})\right)+\frac{1}{2m^2} \left( iH\gamma
t-O(\frac{1}{m}) \right)^2 + \cdots \right)}= e^{-i\gamma H t}. \ee
Here we have used the series expansion of the principal logarithm of
a matrix $\log({\bf 1}-A)=-\sum_{k=1}^\infty \frac{1}{k}A^k$.}
\be{limite} \lim_{x \rightarrow 0} U^{\tau}(x)=\lim_{x\rightarrow
0}\left( e^{-iHx}+O(x^2) \right)^{\frac{\gamma t}{x}}= e^{-i\gamma H
t}. \ee Therefore, over the interval $[0,t]$ the system evolves
according to the Schr\"odinger equation $i\dot \psi=\gamma H \psi$
where $H$ is defined in (\ref{Hamilt}). Defining
$\psi:=[\psi_R^T,\psi_L^T]^T$, with $\psi_R$ and $\psi_L$ both
$N$-dimensional, we have \be{Equo1} \dot \psi_R=-i \gamma ({\bf
1}+F^2)\psi_L, \ee \be{Equo2} \dot \psi_L=-i \gamma ({\bf
1}+F^{2T})\psi_R.  \ee There are special (linear)  combinations of
$\psi_R$ and $\psi_L$ which evolve according to $i\dot
\Psi=\pm\gamma A \Psi$ where $A=F+F^T$ is the adjacency matrix of
the cycle graph. In particular, define \be{psi1piu}
\Psi_{1+}:=\psi_R+F\psi_L. \ee Using (\ref{Equo1}) and (\ref{Equo2})
\be{calcolo} \dot \Psi_{1+}=-i\gamma \left((F^TF+F^2)\psi_L+F({\bf
1}+F^{2T})\psi_R \right)=\ee $$-i\gamma \left( F^T (\psi_R+F
\psi_L)+F(\psi_R+F\psi_L) \right)=-i \gamma A   \Psi_{1+}.
$$
With analogous calculations, after defining \be{psi2piu}
\Psi_{2+}:=\psi_L+F^T \psi_R, \ee \be{psi1meno} \Psi_{1-}:=\psi_R-F
\psi_L, \ee \be{psi2meno} \Psi_{2-}:=\psi_L-F^T \psi_R, \ee one
finds \be{collect} i \dot \Psi_{1,2 \pm}= \pm \gamma A \Psi_{1,2\pm
}. \ee From these equations, it follows that,  defining  \be{fhi}
\Phi_{1,2\pm}:= \frac{e^{\pm 2i \gamma t}}{2}\Psi_{1,2 \pm}, \ee we
have \be{final} i \dot \Phi_{1,2 \pm}=\pm \gamma  L \Phi_{1,2 \pm},
\ee where $L$ is the Laplacian $L=-2 {\bf 1}+A$. These are the
dynamics (forward or backward in time) of the corresponding CTQW.

The dynamics of $[\psi_R,\psi_L]^T$ are obtained as the average of
dynamics of the CTQW corresponding to the adjacency matrix, i.e.,
(from (\ref{psi1piu}), (\ref{psi2piu}), (\ref{psi1meno}),
(\ref{psi2meno})) \be{jklk} \pmatrix{\psi_R \cr \psi_L}=\frac{1}{2}
\left(\pmatrix{ \Psi_{1+} \cr \Psi_{2+}}+ \pmatrix{\Psi_{1-} \cr
\Psi_{2-}} \right). \ee

\section{Extension to general graphs}
\label{extension}

The calculation presented in the previous section follows the  main
steps of \cite{Strauch} which we have adapted to the case of the
cycle. We have however modified this calculation in several
respects. In particular, we have omitted the Fourier transformation
(and anti-transformation) step and replaced the calculation (8) of
\cite{Strauch} which is based on the algebra of Pauli matrices with
a Taylor expansion in a perturbation parameter $x$. In this form,
the treatment  highlights the main ideas of the process to obtain
the dynamics the CTQW as  an appropriate limit of the coined DTQW
and therefore can be generalized. In fact, we will show in this
section how  this limit can be extended to quantum walks on general
graphs and-or CTQW with Hamiltonian different from the Laplacian or
the adjacency matrix. The main idea of the calculation in the
previous section is to take a {\it reference trajectory} of the
coined DTQW which takes the walk back to its original position (up
to an overall phase factor).\footnote{Note that such a trajectory
always exists. As the matrix $S$ in (\ref{defcontrS}) is a
permutation matrix $S^r={\bf 1}$ with $r$ the order of $S$. That is:
by taking $r$ steps with the coin operation $C$ equal to the
identity we come back to the original position of the walk.} Then
one perturbs such a trajectory by slightly modifying the coin
operation at each step. Let $x$ be a parameter which measures the
magnitude  of this perturbation. The resulting trajectory will agree
with $e^{-iHx}$ for a certain {\it simulable Hamiltonian} $H$ up to
higher order terms $O(x^{1+\delta}),$ ($\delta > 0$). Then one
repeats this trajectory a number of times which  increases as $x
\rightarrow 0$ while letting the perturbation go to zero as in
(\ref{limo4}). The result is an evolution of the type $e^{-i \gamma
H t}$ for some scalar $\gamma$. The Hamiltonian $H$ acts on a $c
\times N$ space where $c$ is the dimension of the coin space and $N$
is the dimension of the walker space (i.e., the number of vertices
of the graph) while we would like to obtain an evolution according
to an Hamiltonian $\tilde H$ acting on $N$-dimensional space.
However, if the simulable Hamiltonian $H$ has $N$ eigenvalues
coinciding with the ones of $\tilde H$, by a change of coordinates
we can isolate a subspace of ${\cal C} \otimes {\cal H}$ where the
evolution coincides with the one determined by $\tilde H$. This is
meaning of the calculations in
(\ref{psi1piu})-(\ref{jklk}).\footnote{In this case, in fact, there
are $4$ possible subspaces and, at the limit, the dynamics of the
DTQW gives $4$ copies of the dynamics of CTQW, $2$ forward in time
and  $2$ backward in time.}

\br{evencyc} We remark that the above procedure can be used not only
to obtain  the CTQW as a limit of  the coined DTQW for new graphs
(as we shall see in the following)  but also to obtain alternative
procedures for the case of the cycle. For example, consider the
reference trajectory $S^N={\bf 1}$ with $S$ given in (\ref{esse})
and perturb it as \be{perturbSN} S^N  \rightarrow S e^{Ex} \otimes
{\bf 1} S^{N-1} e^{Ex} \otimes {\bf 1}, \ee with $E:=\pmatrix{0&-i
\cr -i & 0}$. A straightforward calculation shows that we have
\be{conto3} S e^{Ex} \otimes {\bf 1} S^{N-1} e^{Ex} \otimes {\bf
1}={\bf 1}-iHx+O(x^2)=e^{-iHx}+O(x^2), \ee where $H$ is the same as
in (\ref{Hamilt}). Therefore the treatment then goes as after
formula (\ref{U2x}). \er

The question at this point is the characterization of the set of
simulable Hamiltonians for general coined DTQW. To this purpose we
give a more precise definition of simulable Hamiltonian and then
describe this set in Theorem \ref{CARA} below. \bd{SimuHam} An
Hamiltonian $H$ on the Hilbert space ${\cal C} \otimes {\cal V}$
(with $c=\dim{\cal C}$ and $N=\dim{\cal{V}}$) is called simulable if
there exists a parameter $x$ and a sequence of transformations $S
C_j e^{E_j x}\otimes {\bf 1}$, with $C_j \in U(c)$, $E_j \in u(c)$,
$j=1, \ldots, m$, such that \be{product} \prod_{j=1}^m S C_j e^{E_j
f_j(x)}\otimes {\bf 1}= e^{-iHx}+O(x^{1+\delta}), \ee for some
$\delta
>0$, strictly increasing smooth functions,
$f_j$, $f_j(0)=0$ and $x \in [0, \epsilon)$, for some $\epsilon >0$ .\ed

If $H$ is simulable, then one can use a sequence of steps of the
DTQW and perform a limit procedure as in (\ref{limo4}) to obtain the
dynamics corresponding to $H$. In particular, using (\ref{product}),
one obtains \be{LHOP} \lim_{x \rightarrow 0} \left(\prod_{j=1}^m S
C_j e^{E_j f_j(x)}\otimes {\bf 1} \right)^{\frac{\gamma
t}{x}}=\lim_{x \rightarrow 0} \left( e^{-iHx} +O(x^{1+\delta}
\right)^{\frac{\gamma t}{x}}=e^{-iH \gamma t} \ee

\bt{CARA} Let $r$ be the order of the matrix $S$ of a coined DTQW,
 and consider the set \be{calf} {\cal F}=\{ u(c)\otimes {\bf 1}_N, S
u(c)\otimes {\bf 1}_N S^{r-1}, S^2 u(c)\otimes {\bf 1}_N S^{r-2},
\ldots, S^{r-1} u(c)\otimes {\bf 1}_N S \}.\ee Then, the set of
simulable Hamiltonian is (modulo $i$) the Lie algebra generated by
${\cal F}$. \et

We denote by `Sim' the set of simulable Hamiltonians and by ${\cal
L}$ the Lie algebra generated by ${\cal F}$.\footnote{Note this
coincides with the dynamical Lie algebra studied in \cite{DPA} which
determines the set of reachable states for a DTQW on the cycle seen
as a control system.} The statement of the Theorem says that
\be{truT} i \text{Sim}= {\cal L} \ee

\bpr First, we show that $ i \text{Sim} \subseteq  {\cal L}$. Assume
$H \in \text{Sim}$ so that
 (\ref{product}) holds. For every $x$ in $[0,\epsilon)$ the
 left hand side belongs to the Lie group corresponding to the Lie
 algebra $\cal L$, which we denote by $e^{\cal L}$. To see this, we
 first observe that $e^{\cal L}$ has the property

 {\it  Property (a)}:

 $$
X \in e^{\cal L} \Leftrightarrow S^{k} X S^{-k} \in e^{\cal L} ,
 $$
for every integer $k$.

\noindent This property  holds for the Lie algebra ${\cal L}$ (i.e.,
with  $e^{\cal L}$ replaced by ${\cal L}$)  and carries over to the
Lie group $e^{\cal L}$.\footnote{To prove that the property holds
for ${\cal L}$ just observe that it holds for the set ${\cal F}$ in
(\ref{calf}) and it is preserved under linear combination and Lie
bracket of two elements of the Lie algebra.}

Notice that the left hand side of (\ref{product}) (as well as the
right hand side) is equal to the identity when $x=0$. If $m=1$ in
(\ref{product}) then the left hand side is a curve in $e^{\cal L}$
for every $x\in [0,\epsilon)$, $T=T(x)$, with $T(0)={\bf 1}$. This
is true even if $m \geq 2$. To see this, we proceed by induction on
$m$, using {\it Property (a)}. From $\prod_{j=1}^m SC_j \otimes {\bf
1}_N={\bf 1}_{cN}$, we obtain $SC_1 \otimes {\bf 1}=\prod_{j=m}^2
\left( C_j ^\dagger \otimes {\bf 1} S^\dagger \right)$, and
therefore we have \be{poiuy} \prod_{j=1}^m S C_j e^{E_j
f_j(x)}\otimes {\bf 1}=
 \prod_{j=m}^2 \left( C_j
^\dagger \otimes {\bf 1} S^\dagger \right) e^{E_1 f_1(x)} \otimes
{\bf 1} \prod_{j=2}^m S C_j e^{E_j f_j(x)}\otimes {\bf 1}:=L_m(x).
\ee Define $L_1(x)=e^{E_1 f_1(x)} \otimes {\bf 1}$ which is in
$e^{\cal L}$. We have \be{opl}L_{m+1}(x)=C_{m+1}^\dagger \otimes
{\bf 1} S^\dagger L_m(x) S C_{m+1} \otimes {\bf 1} e^{E_{m+1}
f_{m+1} (x)} \otimes {\bf 1}.\ee $S^\dagger L_m(x) S \in e^{\cal L}$
because of {\it Property (a)} and taking the product with element in
$e^{\cal L}$ we obtain an element in $e^{\cal L}$ because of the
group property. In conclusion the left hand side of (\ref{product})
is a smooth curve $T(x)$ in $e^{\cal L}$, with $T(0)={\bf 1}$. From
\be{der89} T(x)=e^{-iHx} +O(x^{1+\delta}),  \ee taking the
derivative with respect to $x$ at $x=0$ we obtain $-iH=\dot T(0)$,
i.e., $-iH$ is an element of the Lie algebra ${\cal L}$. This
conclude the proof that $i\text{Sim} \subseteq {\cal L}$.

To prove that ${\cal L} \subseteq i\text{Sim}$ we prove two facts

\begin{enumerate}

\item
$${\cal F} \subseteq i \text{Sim}$$

\item
$ i \text{Sim} $ is a (real) Lie algebra.\footnote{All the Lie
algebras we are considering are subalgebras of the {\it real} Lie
algebra $u(n)$ (for appropriate $n$).}
\end{enumerate}
From these two facts, since, by definition, ${\cal L}$ is the
smallest subalgebra of $u(cN)$ containing ${\cal F}$, it follows
that ${\cal L} \subseteq i\text{Sim}$. The proof of 1. is almost
obvious. If $-iH \in {\cal F}$ then $e^{-iHx}=S^k e^{Lx} \otimes
{\bf 1} S^{r-k}$ for some $k=1,\ldots,r$, and this can be written in
the form (\ref{product}) by choosing $m=r$, $C_j={\bf 1}$ for every
$j$, $E_j=L$ for $j=k$ and $E_j=0$ otherwise, and $f_j(x)=x$, for
every $j$. To prove 2., we need to prove that: (a) $ iH \in i
\text{Sim} \Leftrightarrow -iH \in i \text{Sim} $; (b) $ iH \in i
\text{Sim} \Rightarrow a iH \in i \text{Sim}$, for each positive
$a$; (c) $ iH_1, i H_2  \in i \text{Sim} \Rightarrow i(H_1 +H_2) \in
i \text{Sim}$; (d) $iH_1, iH_2 \in i \text{Sim} \Rightarrow [iH_1,
iH_2] \in i \text{Sim}$. That is, (a), (b) and (c) show that $i
\text{Sim}$ is a vector space and (d) that is closed under
commutation relation. Rewrite (\ref{product}) as
\be{tobemanipulated} e^{-iHx}= T(x)+O(x^{1+\delta}), \ee (cf.
(\ref{der89})), i.e., by replacing the product in (\ref{product}) by
the symbol $T(x)$, and notice that $T^{-1}(x)$ can be also written
as a product of the form in (\ref{product}). By straightforward
manipulations of (\ref{tobemanipulated}) we obtain
$T^{-1}(x)=e^{iHx}+ T^{-1}(x) O(x^{1+\delta}) e^{iHx}$, that is
$e^{iHx}= T^{-1}(x)+O(x^{1+\delta})$, which shows (a). To show (b),
notice that if $T(x)$ is an admissible product so is $T(ax)$, for $a
>0$, so by replacing $x$ with $ax$ in (\ref{tobemanipulated}), we obtain
the result. To prove (c), denote by $T_1(x)$ and $T_2(x)$ the
product in (\ref{tobemanipulated}) corresponding to $H_1$ and $H_2$,
respectively and $\delta_1$ and $\delta_2$, the corresponding
$\delta$'s in (\ref{tobemanipulated}). From the exponential formula
$e^{-iH_1x}e^{-iH_2x}=e^{-i(H_1+H_2)x}+O(x^2)$, we obtain
$$
e^{-i(H_1+H_2)x}+O(x^2)=\left( T_1(x)+O(x^{1+\delta_1})\right)
\left( T_2(x)+O(x^{1+\delta_2})\right),
$$
which gives
$$
e^{-i(H_1+H_2)x}=T_1(x) T_2(x)+O(x^{1+\min{(1,\delta_1,\delta_2)}}).
$$
which shows  our claim since $T_1 T_2$ is also an admissible
product. Finally, to prove (d) we use the exponential formula
$e^{[-iH_1, -iH_2]t^2}+O(t^3)=e^{-iH_1t}e^{-iH_2t}e^{iH_1
t}e^{iH_2t}$.(see, e.g., \cite{HornJohnsonT}, Section 6.5.).   Using
this and the notation described above, call $O_{1,2}$ the $O$
function for $H_1$ and $H_2$, respectively. We have
$$
e^{[-iH_1, -iH_2]t^2}+O(t^3)= $$
$$\left( T_1(t)+O_1 \right) \left(
T_2(t)+O_2 \right) \left( T_1^{-1}(t)- T_1^{-1}(t)O_1 e^{iH_1
t}\right) \left( T_2^{-1}(t)- T_2^{-1}(t)O_2 e^{iH_2 t}\right).
$$
Expanding the right hand side and omitting explicit terms that are
clearly $O(t^\alpha)$ for $\alpha >2$ since they contain products of
two $O$ functions, we obtain \be{almostdone} e^{[-iH_1,
-iH_2]t^2}+O(t^3)=O(t^\alpha)+ T_1T_2T_1^{-1}T_2^{-1}-
T_1T_2T_1^{-1}T_2^{-1}O_2e^{iH_2t}-T_1T_2T_1^{-1}O_1e^{iH_1t}T_2^{-1}+
\ee $$ T_1O_2T_1^{-1}T_2^{-1}+O_1T_2T_1^{-1}T_2^{-1}. $$ Developing
in a Taylor series the functions that multiply the $O$ functions on
the right hand side of this expression gives cancelations which show
that only terms of the type $O(t^{2+\delta_1})$ and
$O(t^{2+\delta_2})$ possibly remain. In conclusion, we have \be{lpk}
 e^{[-iH_1,
-iH_2]t^2}= T_1(t)T_2(t)T_1^{-1}(t)T_2^{-1}(t)+O(t^\beta), \ee with
$\beta >2$ and by setting $x=t^2$, we obtain \be{lpkuuuu}
 e^{[-iH_1,
-iH_2]x}=
T_1(\sqrt{x})T_2(\sqrt{x})T_1^{-1}(\sqrt{x})T_2^{-1}(\sqrt{x})+O(x^{\frac{\beta}{2}}),
\ee which show the result because if $T(x)$ is an admissible product
so is $T(\sqrt{x})$. This concludes the proof of the theorem.
 \epr

\section{Example}
\label{Example}

\begin{figure}[ht]

\begin{picture}(340,200)(0,22)

\linethickness{0.6pt}

%%vertici
\put(168,144){{\Large $\bigcirc$}} \put(169,144){{$A$}}

\put(268,144){{\Large $\bigcirc$}}\put(269,144){{ $C$}}

\put(168,44){{\Large $\bigcirc$}} \put(169,44){{ $B$}}

\put(268,44){{\Large $\bigcirc$}} \put(269,44){{ $D$}}

%%archi
\put(184,146){\line(1,0){84}}

\put(184,46){\line(1,0){84}}

\put(177,139){\line(0,-1){84}}

\put(273,139){\line(0,-1){84}}

\put(182,140){\line(1,-1){88}}

\put(271,142){\line(-1,-1){88}}

%labels

\put(223,150){{$1$}}

\put(223,50){{$1$}}

\put(165,100){{$2$}}

\put(276,100){{$2$}}

\put(186,120){{$3$}}

\put(242,120){{$3$}}
%\put(179,130){{\footnotesize $2$}}

\end{picture}
\caption{Graph for the  discrete time quantum walk in the example.
The labels on the edges indicate the corresponding coin toss result.
In this example the same coin toss result induces motion in both
directions. }. \label{F1}
\end{figure}
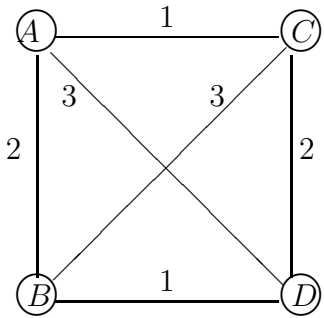

I consider  now an example of a coined DTQW whose associated CTQW
can be obtained as an appropriate  limit. This example is not a DTQW
on cycle or $d$-lattice and shows the generality of the method
described above. In particular, consider the graph in Figure
\ref{F1} which has a three dimensional coin space ${\cal C}$
(spanned by the coin tossing results $1,2,3$) and a $4$-dimensional
walker space spanned by the positions $A,B,C,D$, in that order.

The associated matrix $S$ is given by $S=\text{diag}(S_1,S_2,S_3)$,
with \be{esse123} S_1=\pmatrix{0 & 0 & 1 & 0\cr 0& 0 & 0 & 1 \cr 1 &
0 & 0 & 0\cr 0 & 1 & 0 & 0}, \qquad S_2=\pmatrix{0 & 1 & 0 & 0 \cr
1& 0 & 0 & 0 \cr 0 & 0 & 0 & 1\cr 0 & 0 & 1 & 0}, \quad
S_3=\pmatrix{0 & 0 & 0 & 1 \cr 0& 0 & 1 & 0 \cr 0 & 1 & 0 & 0\cr 1 &
0 & 0 & 0}.\ee The Lie algebra of simulable Hamiltonians $i
\text{Sim}$ is generated by \be{hju} {\cal F}= \{ u(3) \otimes {\bf
1}_4, S  u(3) \otimes {\bf 1}_4 S \}, \ee since $S$ has order $2$.
It contains, in particular, the element $\text{diag} (-3i, i, 2i)
\otimes {\bf 1}_4$. The adjacency matrix $A:=S_1+S_2+S_3$ has one
eigenvalue equal to $3$ with multiplicity $1$ and one eigenvalue
$-1$ with multiplicity $3$. Therefore, we can single out  a
$4$-dimensional vector space of ${\cal C} \otimes {\cal V}$ of the
DTQW where the dynamics coincide with the one of the CTQW
corresponding to the adjacency matrix. We can, in fact, do better by
studying more closely the structure of the Lie algebra generated by
${\cal F}$ in (\ref{hju}). $S_1$, $S_2$ and $S_3$ can be
simultaneously diagonalized, and moreover we have $S_1 S_2=S_3$,
$S_2 S_3=S_1$ and $S_3 S_1=S_2$ and $S_1^2=S_2=S_3^2={\bf 1}_4$. The
set ${\cal F}$ can be written as \be{calf2} {\cal F}=\left \{ u(3)
\otimes {\bf 1}_4, E_{12} \otimes S_{3}, E_{13} \otimes S_2, E_{23}
\otimes S_1,  \right\}\ee where $E_{jk}$ represents the set of all
the matrices in $su(3)$ with all the entries equal to zero except
for the ${j,k}$ and ${j,k}$ entries. From this, one finds that
\be{clL} {\cal L}= \text{span} \left\{
    u(3)\otimes {\bf 1}_4, su(3) \otimes {S_1}, su(3) \otimes {S_2}, su(3) \otimes {S_3}. \right\}\ee
It is in fact clear that this is a Lie algebra containing ${\cal
F}$. Moreover by calculating the set $\bigoplus_{k=0}^\infty
ad_{su(3)}^k E_{jk}$, for a given $jk$ it is clear that this is all
of $su(3)$, since it is a non-empty ideal in $su(3)$, which is a
simple Lie algebra. Therefore, we obtain with repeated Lie brackets
all the  matrices in $su(3)\otimes S_{1,2,3}.$ At this point, it is
convenient to make a change of coordinates to simultaneously
diagonalize $S_1,$ $S_2$ and $S_3$ and to change the order of the
tensor product in (\ref{clL}). Therefore the matrices in ${\cal L}$
are linear combinations of matrices of the form
$L_1:=\text{diag}(A,A,A,A)$, $L_2:=\text{diag}(B,B,-B,-B)$,
$L_3:=\text{diag}(-C,C,-C,C)$, $L_4:=\text{diag}(-D,D,D,-D)$, with
$A\in u(3)$, and $ B,C,D \in su(3)$. We can choose a linear
combination of $L_{1,2,3,4}$ to make the resulting matrix  have
eigenvalues $\pm 3i$ with multiplicity $1$ and $\pm i$ with
multiplicity $3$ and $0$ with multiplicity $4$, so as to obtain 2
copies of the CTQW (one backward and one forward in time) and no
dynamics on appropriate four dimensional subspaces of ${\cal C}
\otimes {\cal V}$.

\section{Conclusions}
\label{conclu}

I have obtained the dynamics of the continuous time quantum walk on
the cycle as an appropriate limit  of the coined discrete time
quantum walk on the same graph. This result can be extended directly
to the case of $d$-lattices using the fact that they are products of
cycles. The main ideas of the   procedure can be extended to general
graphs. One can use the discrete time quantum walks to obtain the
dynamics corresponding to a given Hamiltonian on the whole
(coin+walker) space and then obtain the dynamics of the continuous
time quantum walk by restricting oneself to a subspace of the whole
(coin+walker) space. I have characterized the set of Hamiltonians
for which this is possible, a set that I called of simulable
Hamiltonians. The results of this paper reduce the problem of
obtaining the continuous time quantum walk from the discrete time to
the problem to a study of the set of simulable Hamiltonians.  This
set has the structure of a Lie algebra. As this set is quite rich it
is reasonable to expect that every (or at least most) continuous
time quantum walks can be obtained by a limiting procedure and
restricting to an appropriate subspace the dynamics of the
associated discrete time quantum walk.

%1) Two types of generalizations

%2)Open problem: is the adjacency matrix always simulable?

\end{document}